\documentclass[aps,prd,twocolumn,nofootinbib]{revtex4-1}
\usepackage{epsfig}
\usepackage[colorlinks,linkcolor=blue,anchorcolor=blue,citecolor=blue,urlcolor=blue,breaklinks=true]{hyperref}
\usepackage{graphics}
\usepackage{slashed}
\usepackage{color}
\usepackage{amsfonts}
\usepackage{amsmath}
\usepackage{extarrows}
\usepackage{amssymb}
\usepackage{epstopdf}
\usepackage{float}
\usepackage{multirow}

\begin{document}
\title{Investigation of exotic state X(3872) in $pp$ collisions at $\sqrt{s}=7$, 13~TeV}
\author{Hong-ge Xu$^{1,2}$}
\author{Zhi-Lei She$^{1,2}$}
\author{Dai-Mei Zhou$^{3}$}
\author{Liang Zheng$^{2,3}$}
\author{Xiao-Lin Kang$^{2}$}
\author{Gang Chen$^{2}$}\email{Email:chengang1@cug.edu.cn(Corresponding Author)}
\author{Ben-Hao Sa$^4$}

\affiliation{
${^1}$Institute of Geophysics and Geomatics, China University of Geosciences, Wuhan 430074, China\\
${^2}$School of Mathematics and Physics, China University of Geoscience, Wuhan 430074, China\\
${^3}$Key Laboratory of Quark and Lepton Physics, Central China Normal University, Wuhan 430079,China\\
${^4}$China Institute of Atomic Energy, P.O. Box 275(10), Beijing 102413, China}

\begin{abstract}
We have used the dynamically constrained phase space coalescence model to study the production of the exotic state $X(3872)$ based on the hadronic final states generated by the parton and hadron cascade model (PACIAE) with $|y|<1$ and $p_T < 15.5$ GeV/c in $pp$ collisions at $\sqrt{s}=7$ and 13 TeV, respectively. Here the $X(3872)$ is assumed to consist of bound state $D\bar {D^*}$, which can form three possible structures for the tetraquark state, the nucleus-like state, and the molecular state. The yields of three different structures $X(3872)$ were predicted. The transverse momentum distribution and the rapidity distribution of three different structures $X(3872)$ are also presented. Sizable difference can be found in the transverse momentum and rapidity distributions for the three different $X(3872)$ structures.

\end{abstract}

\maketitle

\section{Introduction}
Hadrons are composite objects formed due to strong interactions formulated by the Quantum Chromodynamics (QCD) theory within the quark model framework. The quark model was formulated in 1964 to classify mesons as bound states made of a quark-antiquark pair, and baryons as bound states made of three quarks. For a long time all known mesons and baryons can be classified within this scheme. However, in principle QCD allows the existence of exotic states: multiquark states, hybrid states and glueball, as was already recognized by Gell-Mann in one of the first publications on the quark model~\cite{RefA1}, like $\bar{q}\bar{q}qq$, $\bar{q}\bar{q}\bar{q}qqq$ $\bar{q}qqqq$ and so on. We will refer to any state which does not appear to fit with the expectations for an ordinary $\bar{q}q$ or $qqq$ hadron in the quark model as ``exotic". Due to the non-perturbative properties of QCD in the low energy region, it is unavailable for us to use it to study the hadron structures and the hadron-hadron interactions directly. The study of exotic states can provide essential information on low energy QCD, which is absent in the ordinary qqq baryons and $\bar{q}q$ mesons. Therefore, comprehensive efforts have been made to test their existence and measure their properties.

The first exotic state $X(3872)$ was discovered by the Belle collaboration almost two decades ago in the decay channel $B^{\pm}\rightarrow k^{\pm}\pi^{+}\pi^{-}J/\psi$, with a significance of 10$\sigma$~\cite{RefA2}. The new hadron was then confirmed by CDF~\cite{RefA3}, D0~\cite{RefA4}, BABAR~\cite{RefA5}, and recently also by LHCb~\cite{RefA6}. Forthermore, Belle and other collaborations have reported a lot of ``XYZ" particles ~\cite{RefA7,RefA8,RefA9}, which stimulated many researches on hadron spectrum. Some of the new states are unambiguously interpreted as conventional $c\bar{c}$ states, some are manifestly exotic, while for the others a definite interpretation is still missing. Despite the large amount of experimental data, the nature of the $X(3872)$ state is still unclear. Several interpretations have been proposed, such as the conventional $\chi_{c1}(2P)$ state~\cite{RefB41}, the molecular state~\cite{RefB42,RefB43,RefB44}, the tetraquark state~\cite{RefB45}, the $c\bar c g$ hybrid state~\cite{RefB46}, the vector glueball~\cite{RefB47} or the mixed state~\cite{RefB48,RefB49}.

In recent years, various theoretical models have been proposed to explain these exotic states. Meanwhile, these efforts have been made with various theoretical approaches like effective field theory~\cite{Ref-zheng1}, QCD sum rule~\cite{Ref-zheng2}, lattice QCD~\cite{Ref-zheng3}, gauge invariant model~\cite{Ref-zheng4}, potential model~\cite{Ref-zheng5}. The extreme proximity of the $X(3872)$ to the $D\bar{D}^{*}$ threshold suggests that it could be a $D^{*0}\bar{D}^{0}$, $D^{0}\bar{D}^{*0}$, $D^{-}{D}^{*+}$ or $D^{+}{D}^{*-}$ (hadron molecule or binging state).

Exploring the nature of exotic multiquark candidates such as the $X(3872)$ plays a pivotal role in understanding quantum chromodynamics (QCD). In present days the study of exotic hadron has focused on the decay process, but the production process could actually provide more information. Despite significant efforts, there is still a lack of consensus on the process by which it came into being and its internal structure. There are indications that the multi-quark states may directly produced from multiproduction process of high-energy collisions, rather than from hadron decay~\cite{RefD37p}. Here taking $X(3872)$ as an example, the parton and hadron cascade model (PACIAE)~\cite{RefD37} plus the dynamically constrained phase space coalescence model (DCPC)~\cite{RefD34,RefD35,RefD36} is proposed to simulate the complete evolution process from the initial state of the parton to the final state of the multiparticle in high-energy collisions. Therefore, we predict the yields and properties of the exotic hadron $X(3872)$ and explore the production process and its structure of the exotic hadron $X(3872)$ in high-energy pp collisions. This would provided a new and effective method for us to understand and study exotic hadrons.

In this paper, we will mainly consider three different structures of $X(3872)$, tetraquark state, nucleus-like state, and molecular state. An approach for DCPC model is introduced to connect the multi-particle final state with the PACIAE simulations and to calculate the production of exotic state $X(3872)$ in ultrarelativistic $pp$ collisions. Firstly, the PACIAE models were used to generate the hadronic final states, including $D^{*0}$, $\bar D^{0}$, $D^{0}$, $\bar D^{*0}$, $D^{-}$, $D^{*+}$, $D^{+}$, and $D^{*-}$. Then the yields, the transverse momentum distribution, as well as the rapidity distribution for $X(3872)$ as tetraquark state, nucleus-like state, and molecular state were predicted using DCPC model in $pp$ collisions at $\sqrt{s}=7$ and 13 TeV.

\section{{\footnotesize PACIAE} Model and {\footnotesize DCPC} Model}
The parton and hadron cascade model {\footnotesize PACIAE}~\cite{RefD37} is based on {\footnotesize PYTHIA} 6.4 to simulate various collision, such as $e^{+}e^{-}$, $pp$, $p$-$A$ and $A$-$A$ collisions. In general, {\footnotesize PACIAE} has four main physics stages, consisting of the parton initiation, parton rescattering, hadronization, and hadron rescattering. In the parton initiation, the string fragmentation is switched off temporarily in {\footnotesize PACIAE} and di(anti)quarks are broken into (anti)quarks. This partonic initial state can be regarded as quark-gluon matter (QGM) formed inside the parton initialization stage in the $pp$ collisions. Then the parton rescattering in QGM is taken into account by the $2\to 2$ LO-pQCD parton-parton cross sections\cite{PYTHIA6.4}. Their total and differential cross sections in the parton evolution are computed by the Monte Carlo method. In the hadronization process, the parton can be hadronized by the Lund string fragmentation re gime and/or the phenomenological coalescence model~\cite{PYTHIA6.4}. The final stage is the hadron rescattering process happening between the created hadrons until the hadronic freeze-out.

Theoretically, the production of composite hadronic states like light nuclei are usually calculated in two steps: First, the primordial simple hadrons including mesons and baryons are obtained with the transport model approach. Then, the nuclei or bound states are calculated by the phase-space coalescence model based on the Wigner function~\cite{New45,New452} or by the statistical model~\cite{New46}. We introduced a dynamically constrained phase-space coalescence ({\footnotesize DCPC}) model to calculate the yield of bound states after the transport model simulations.

From quantum statistical mechanics~\cite{reff1}, one can not simultaneously define both position $\vec q\equiv (x,y,z)$ and momentum $\vec p \equiv (p_x,p_y,p_z)$ of a particle in six-dimensional phase space because of the uncertainty principle, $\Delta \vec q\Delta \vec p \sim h^3$. One can only say this particle lies somewhere within a six-dimensional quantum box or state of volume of $\Delta \vec q\Delta \vec p$ volume element in the six-dimensional phase space corresponding to a state of the particle. Therefore, one can estimate the yield of a single particle~\cite{reff1} by

\begin{equation}
Y_1=\int_{E_{a}\leq H\leq E_{b}} \frac{d\vec qd\vec p}{h^3},
\end{equation}
where $E_{a}, E_{b}$, and $H$ denote energy threshold and the energy function of the particle, respectively. The variables $\vec q$ and $\vec p$ are the coordinates and momentum of the particle in the center-of-mass frame of the collision at the moment after hadronization. Furthermore, the yield of a cluster consisting of $N$ particles is defined as following:
\begin{equation}
Y_{N} =\int\cdots\int_{E_{a}\leq H\leq E_{b}}{\frac{d\vec q_{1}d\vec p_{1}\cdots d\vec q_{N}d\vec p_{N}}{(h)^{3N}}}.
\end{equation}

Therefore, the yield of a $X(3872)$ consisting of $D\bar D^{*}$ cluster in the {\footnotesize DCPC} model can be calculated by

\begin{align}
Y_{X(3872)} =&\int ...\int\delta_{12}\frac{d\vec q_1d\vec p_1 d\vec q_2d\vec p_2}{h^{6}},
\label{yield} \\
\delta_{12}=&\left\{
  \begin{array}{ll}
  1 \hspace{0.2cm} \textrm{if} \hspace{0.2cm} 1\equiv D, 2\equiv \bar D^{*};\\
    \hspace{0.3cm} m_{X(3872)}-\Delta m\leq m_{inv}\\
     \hspace{0.5cm}\leq m_{X(3872)}+\Delta m; \\
    \hspace{0.3cm} q_{12}\leq D_{0};\\
  0 \hspace{0.2cm}\textrm{otherwise}.
  \end{array}
  \right.
\label{yield1}
\end{align}

\textrm{where},
\begin{equation}
\hspace{0.5cm}  m_{inv}=\sqrt{(E_1+E_2)^2-(\vec p_1+\vec p_2)^2}.
\label{yield2}
\end{equation}
 The $|\vec{q}_{12}|=|\vec{q}_{1}-\vec{q}_{2}|$ represent the distance between $D$ and  $\bar D^{*}$, $m_{X(3872)}$ denotes the rest mass of $X(3872)$, and $\Delta m$ refers to its mass uncertainty. $E_1$, $E_2$ and $p_{1}$, $p_2$ denote the energies and momenta of the two particles ($D$ and $\bar D^{*}$), respectively. Here, the $X(3872)$ is produced by the combination of hadrons $D$ and $\bar D^{*}$ after the mult-particle final states have been produced using the PACIAE model in $pp$ collisions at $\sqrt{s}=7$ and 13 TeV. According to the different distances $q_{12}$ between $D$ and $\bar D^{*}$, the exotic state $X(3872)$ can be separated into three structures, the tetraquark state as $D_0<0.6$~fm, the nuclear-like state as $0.6<D_0<1.36$~fm and the molecular state as $1.36<D_0<7$~fm~\cite{radius1,radius2,radius3,radius4}.

\section{Results}
In this work, the final states hadrons are produces by the PACIAE model. The model parameters of PACIAE were fixed on the default values from the PACIAE model, except the parameters of parj(1), parj(2), parj(3), and parj(4). Here, parj(1) is the suppression of diquark-antidiquark pair production compared with the quark-antiquark pair production, parj(2) is the suppression of strange quark pair production compared with up (down) quark pair production, parj(3) is the extra suppression of strange diquark production compared with the normal suppression of a strange quark, and parj(4) is the suppression of spin 1 diquarks compared with spin 0 ones excluding the factor 3 coming from spin counting. These parameters are determined by fitting to the  ALICE data~\cite{D-mesons,pion-kion} of $D^{0}$, $D^{+}$, $D^{*+}$, $\pi^{+}$ and $K^{+}$ in mid-rapidity $pp$ collisions at $\sqrt{s}=7$ TeV. And the comparison of the yields for each final states between the experimental measurement and the calculation from PACIAE model are shown in Table I, which are consistent with each other within uncertainty. The parameters parj(1), parj(2), parj(3 )and parj(4) found to be 0.08, 0.45, 0.40, 0.36, respectively. Notice that the yield of  $D^{0}$, $D^{+}$ and $D^{*+}$ for the experimental data in Table I is calculated according to the cross sections in Refs~\cite{D-mesons}.

\small\addtolength{\tabcolsep}{1.pt}
\begin{table}[ht]
\centering
\caption{The yields of $D^{0}$, $D^{+}$, $D^{*+}$, $\pi^{+}$ and $K^{+}$ computed by PACIAE model in mid-rapidity $pp$ collisions at $\sqrt{s}=7$ TeV and comparison with experimental data~\cite{D-mesons,pion-kion}, with the $0<p_{T}<36$~GeV/c for $D^{0}$, $1<p_{T}<24$~GeV/c for $D^{+}$ and $D^{*+}$, $0.1<p_{T}<3$~GeV/c for $\pi^{+}$, and $0.2<p_{T}<6$~GeV/c for $K^{+}$.}
\setlength{\tabcolsep}{11.5pt}
\renewcommand{\arraystretch}{1.2}
\begin{tabular}{cccc} \hline  \hline
 particles &Experiment &PACIAE  \\ \hline
 $D^{0}$ &$(8.04\pm1.57)\times10^{-3}$ &$9.55\times10^{-3}$ \\
 $D^{+}$  &$(2.93\pm0.72)\times10^{-3}$ &$2.67\times10^{-3}$ \\
 $D^{*+}$  &$(3.33\pm0.86)\times10^{-3}$ &$3.88\times10^{-3}$ \\
 $\pi^{+}$ &$2.26\pm0.10$ &2.17 \\
 $K^{+}$  &$0.286\pm0.016$ &0.280 \\ \hline \hline
\end{tabular} \label{paci1}
\end{table}

\begin{figure*}[!htb]
\includegraphics[width=0.9\textwidth]{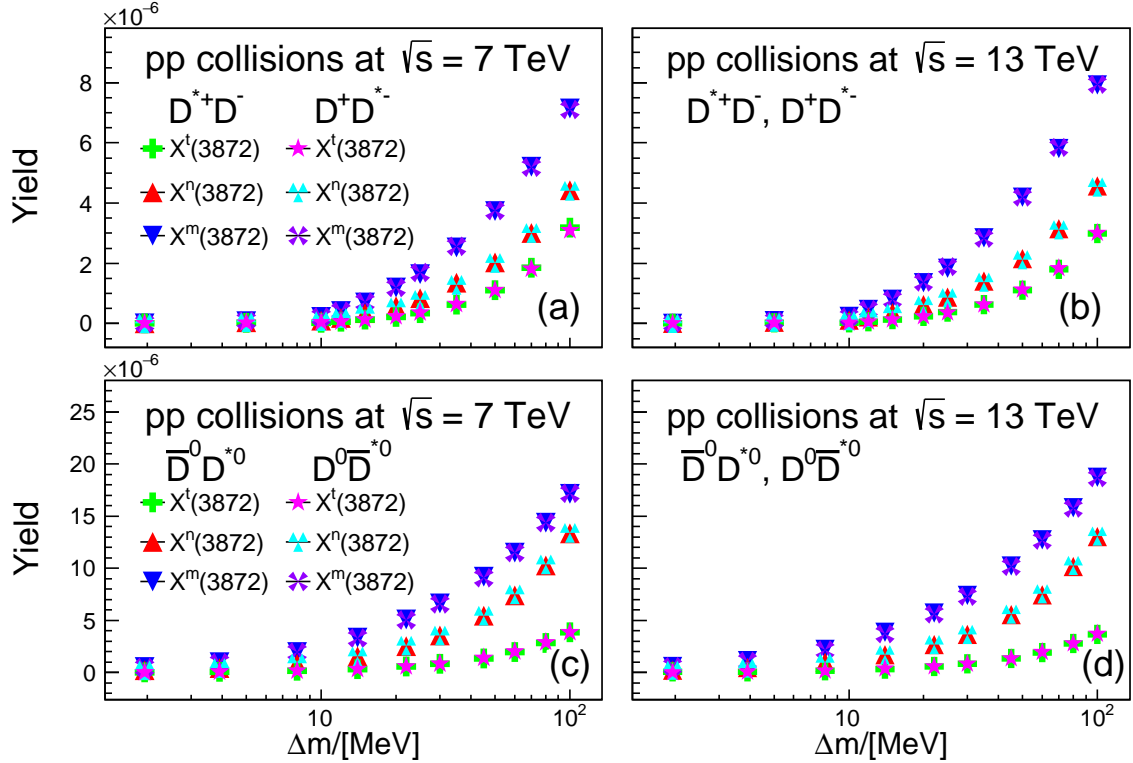}
\caption{The distribution of the yield of exotic states $X(3872)$ in $pp$ collisions with $|y|< 1$ and $0<p_{T}<15.5$ GeV/c, as a function of mass uncertainty $\Delta m$. The data are calculated using PACIAE+DCPC model as $X(3872)$ consists of the bound state $D\bar{D}^{*}$, (a) as $X(3872)$ consists of the bound states ${D}^{*+}{D}^-$ and $D^+{D}^{*-}$ at $\sqrt{s}=7$~TeV, (b) as $X(3872)$ consists of the bound states $D^+{\bar D}^{*-}$ and ${\bar D}^-{D}^{*+}$ at $\sqrt{s}=13$~TeV, (c) as $X(3872)$ consists of the bound state $\bar {D}^0{D}^{*0}$ and $D^0\bar{D}^{*0}$ at $\sqrt{s}=7$~TeV, (d) as $X(3872)$ consists of the bound states $\bar {D}^0{D}^{*0}$ and $D^0\bar{D}^{*0}$ at $\sqrt{s}=13$~TeV. In the figure, symbols with different shapes and different colors are used to represent different exotic states $X(3872)$ constructed by different bound states $D\bar D^*$.}
\label{tu1}
\end{figure*}
\begin{large}
\small\addtolength{\tabcolsep}{1.pt}
\begin{table*}[!htb]
\caption{The yield of $X(3872)$ of different structures produced by $D\bar{D^*}$ clusters in $pp$ collision at $\sqrt{s}=7, 13$ TeV with $|y|<1, 0<p_{T}<15.5$ GeV/c, computed by the PACIAE+DCPC model, where $\Delta m=\Gamma/2$.}
\setlength{\tabcolsep}{4.5mm}
\begin{tabular}{ccccccc} \hline  \hline
Class of  &\multicolumn{3}{c} {$\sqrt s=7$~TeV} &\multicolumn{3}{c} {$\sqrt s=13$~TeV} \\ \cline{2-4}\cline{5-7}
bound state &$X^{t}(3872)$ &$X^{n}(3872)$& $X^{m}(3872)$&$X^{t}(3872)$ &$X^{n}(3872)$ & $X^{m}(3872)$ \\ \hline
$D^{*0}\bar{D}^{0}$  &3.83$\times10^{-8}$   &2.00$\times10^{-7}$  &5.11$\times10^{-7}$  &4.60$\times10^{-8}$  &2.39$\times10^{-7}$  &5.89$\times10^{-7}$  \\
$D^{0}\bar{D}^{*0}$  &3.87$\times10^{-8}$   &2.08$\times10^{-7}$  &4.68$\times10^{-7}$  &3.87$\times10^{-8}$  &2.19$\times10^{-7}$  &5.57$\times10^{-7}$  \\
$D^{-}\bar{D}^{*+}$  &2.52$\times10^{-9}$   &6.81$\times10^{-9}$  &1.37$\times10^{-8}$  &2.72$\times10^{-9}$  &7.18$\times10^{-9}$  &1.55$\times10^{-8}$  \\
$D^{+}\bar{D}^{*-}$  &2.70$\times10^{-9}$   &6.28$\times10^{-9}$  &1.41$\times10^{-8}$  &2.57$\times10^{-9}$  &6.97$\times10^{-9}$  &1.56$\times10^{-8}$  \\
Total.&8.22$\times10^{-8}$&4.21$\times10^{-7}$ &1.01$\times10^{-6}$& 9.0$\times10^{-8}$ &4.72$\times10^{-7}$ & 1.18$\times10^{-6}$ \\
\hline
\end{tabular} \label{paci2}
\end{table*}\end{large}

Three billion $pp$ collision events are generated with the PACIAE model with $|y|< 1$ and $0<p_T<20$ GeV/c at $\sqrt{s}=7$ and 13 TeV, respectively. Then the final state particles of $\bar D^{0}$, $D^{0}$, $\bar D^{*0}$, $D^{*0}$, $D^{+}$, $D^{*+}$, $D^{-}$ and $D^{*-}$ are put into the DCPC model to compute the exotic state $X(3872)$ clustering by $D^{*0}\bar{D}^{0}$, $D^{0}\bar{D}^{*0}$, $D^{-}{D}^{*+}$ and $D^{+}{D}^{*-}$ according the Eqs.(3) and (4). The radius (the distance between the two mesons) and the mass value of $X(3872)$ are given. Here we assume the excited states $X(3872)$ is made up of $D\bar{D}^{*}$ bound state, which is produced during the hadronization or hadronic evolution period.

There are three popular configurations for the production of exotic state $X(3872)$ in multiple production processes: the hadronic molecular state, the nuclear-like states, or the compact tetraquarks. We can determine the structure type of $X(3872)$ from its radius value, and define the radii of the three structures as 8 fm, 1.36 fm and 0.6 fm~\cite{radius1,radius2}, respectively. When the distance $D_0$ between two $D$ mesons is greater than 1.36 fm and less than 8 fm, the molecular state of $X(3872)$ consists of two mesons $D\bar {D}^*$~\cite{radius2}, denided as $X^m(3872)$; When $D_0 \le 0.6$ fm, four quarks $[ c\bar c q\bar q]$ in $X(3872)$ form a compact tetraquarks~\cite{radius1} through strong interaction, denote as $X^t(3872)$; When 0.6 fm $\le D_0 \le 1.36$ fm, $X(3872)$ is considered to consist of two mesons $D\bar{D}^*$ to form a nuclear-like states, denote as $X^n(3872)$. The yield distribution of the three exotic states $X(3872)$ with different structures to the uncertain mass parameter $\Delta m$ is shown in Fig. 1.


As shown in Fig.1, the yields of the $X(3872$) increase rapidly with parameter $\Delta m$ from 0 to 100 MeV. At the same $\Delta m$ and c.m. energy, the yields of molecular state $X^m(3872)$ are greater than the nucleus-like state $X^n(3872)$, and the yield of the nucleus-like state $X^n(3872)$ are greater than that of the tetraquark state $X^t(3872)$.
\begin{figure*}[!htb]
\includegraphics[width=0.83\textwidth]{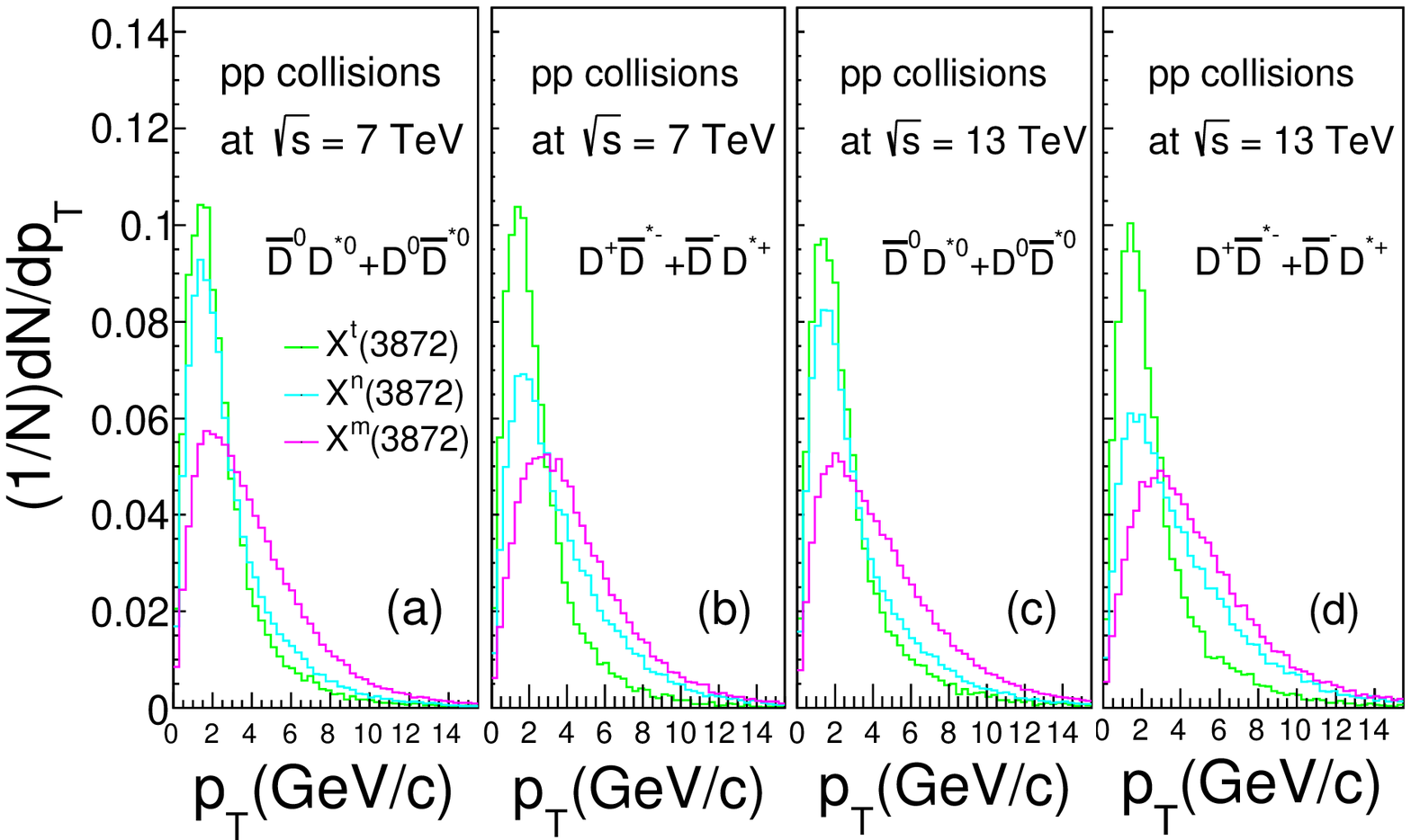}\\
\caption{Transverse momentum distributions of $X(3872)$ of different structures produced for different $D\bar{D^*}$ clusters in $pp$ collision at $\sqrt{s}=7$, 13 TeV  with $|y|<1$, $0<p_{T}<15.5$ GeV/c, computed by the PACIAE+DCPC model.}
\label{tu2}
\end{figure*}

\begin{figure*}[!htb]
\includegraphics[width=0.83\textwidth]{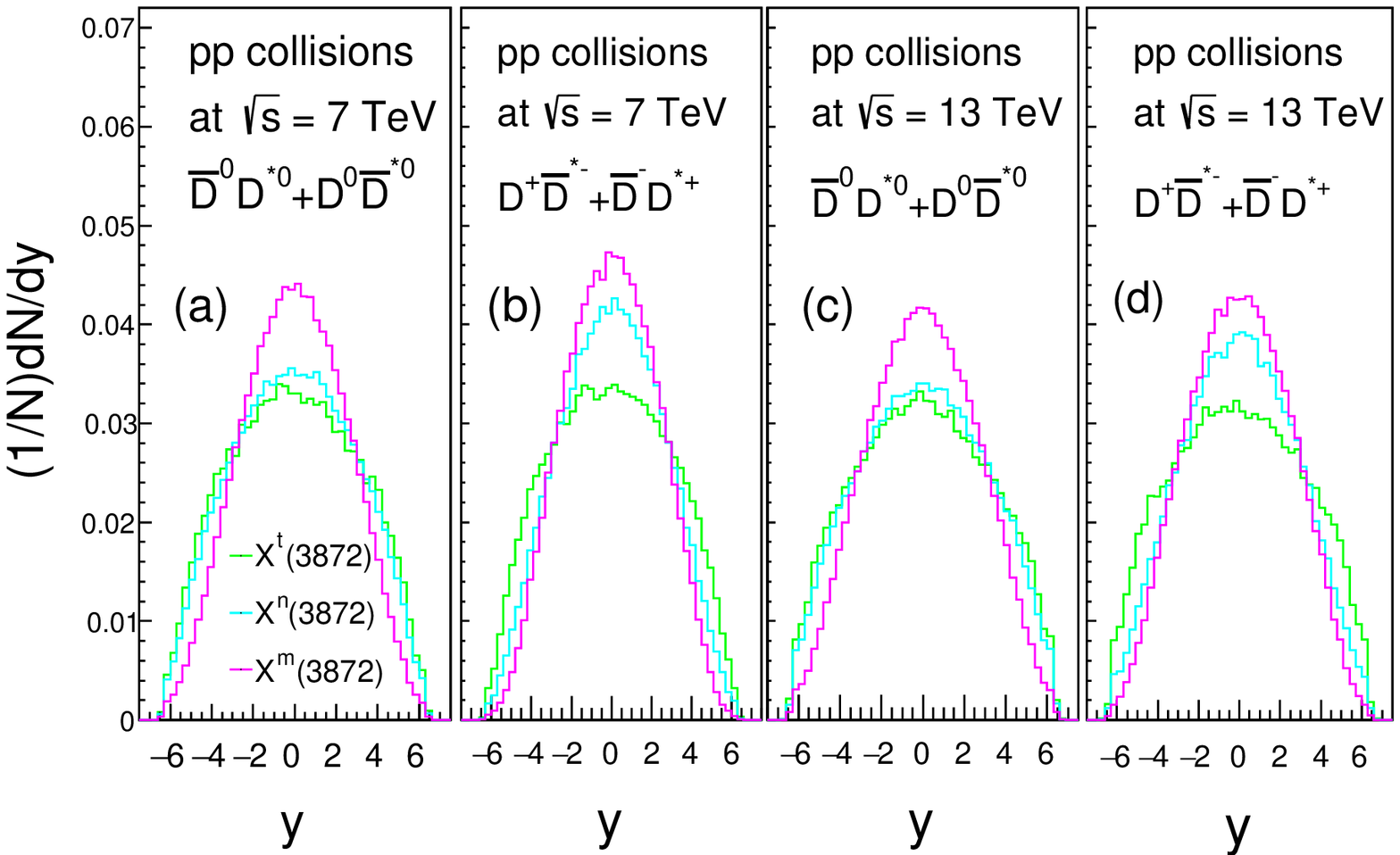}
\caption{Similar to Fig. 2, but for the rapidity distribution. }
\label{tu3}
\end{figure*}

If we take half the decay width of $X(3872)$ as the $\Delta m$ parameter, i.e, $\Delta m=\Gamma/2$, then the yield of the $X(3872)$ would be predicted. When we take the Breit-Wigner width of the $X(3872)$ state that is measured to be ($\Gamma =3.9^{+2.8+0.2}_{-1.2-1.1}$ MeV)~\cite{X3872-Bell2010} by the experiment, ie. $\Delta m=\Gamma/2=1.95$~MeV, the yield of exotic states $X(3872)$ from different decay channels and different structures is given by PACIE+DCPC model in $pp$ collision at $\sqrt{s}=7$ and 13 TeV, as shown in Table II.

Results from Table II show that the yields of the $X(3872$) from $ D^{*0}\bar{D}^{0}$ and $D^{0}\bar{D}^{*0}$ clusters are two orders of magnitude higher than the yields of $ D^{-}{D}^{*+}$ and $ D^{+}{D}^{*-}$ clusters. This indicates that the yield of $X(3872)$ produced in $pp$ collisions mainly comes from the $D^{*0}\bar{D}^{0}$ and $D^{0}\bar{D}^{*0}$ clusters. While the yields of the $X(3872$) from $D^{*0}\bar{D}^{0}$ and $D^{0}\bar{D}^{*0}$ clusters are almost same to each other and the yield of the $X(3872$) from $ D^{-}{D}^{*+}$ and $ D^{+}{D}^{*-}$ clusters are also close. In the last row of Table II, the total yields of the three different structured $X(3872$) particles produced in the $pp$ collision are given. Obviously, the yield of $X^m(3872)$ $(\sim 10^{-6})$ is greater than that of $X^n(3872)$ $(\sim 10^{-7})$, and the yield of $X^n(3872)$ $(\sim 10^{-7})$ is greater than that of $X^t(3872) $ $(\sim 10^{-8})$, that is, the yield of $X^m(3872)$ is the highest in high-energy $pp$ collisions. In addition, as c.m energy increases from 7 to 13 TeV, the yields of the $X(3872)$ significant increased. The ratio between the different $X(3872)$ structures are shown in Table III.

In fact, the generation and decay of exotic hadron $X(3872)$ had been studied in the non-relativistic wave function method in reference~\cite{ref-Jin}. Employing the factorized formulation with the help of event generators, they investigate the production of exotic hadrons in multi-production processes at high energy colliders. The distributions of rapidity and transverse momentum of $X(3872)$ obtained by our simulation computed are close to their results. In reference~\cite{ref-Guo}, Pythia and Herwig models were used to simulate the production of $X(3872)$ as a hadron molecule in the proton-proton/antiproton colliders, and the yields were estimated with the same order of magnitude as those calculated by us.

\begin{large}
\small\addtolength{\tabcolsep}{1.pt}
\begin{table}[!htb]
\caption{The yield ratio of $X(3872)$ of different structures produced in $pp$ collision at $\sqrt{s}=7$ and 13 TeV. }
\begin{tabular}{cccccc} \hline  \hline
 Ratio &$\sqrt{s}=7$~TeV&$\sqrt{s}=13$~TeV   & \\ \hline
$X^{m}(3872)$/$X^{t}(3872)$ &12.29 &13.11 & \\
$X^{m}(3872)$/$X^{n}(3872)$ &2.40 &2.50 & \\
$X^{n}(3872)$/$X^{t}(3872)$&5.12 &5.24 \\
\hline  \hline
\end{tabular} \label{paci3}
\end{table}\end{large}

It should be noted that there are some uncertain factors in the yield of $X(3872)$ given by us. First of all, the determination of the radius parameter for the $X(3872)$ of three structures refers to the results of some theoretical calculations~\cite{radius1,radius2} rather than the experimental measurement of the radius of $X(3872)$. The selection of different radius values will lead to different yield values. Secondly, we take the half-decay width $\Gamma/2$ of the mass spectrum for $X(3872) \to D^{0*}\bar{D^0} + c.c.$ as the mass uncertainty parameter, and the half-decay width ($\Gamma =3.9^{+2.8+0.2}_{-1.2-1.1}$ MeV)~\cite{X3872-Bell2010} measured in the experiment has a large systematic error, which also increases the uncertainty of $X(3872)$yield.

Since this is direct production, and only part of the $X(3872)$ can decay again to $D$'s via strong interaction, this will affect the production ratio of the $D$'s~\cite{ref-Han}. In fact, the measurements of the $X(3872)$ decays including $X(3872) \to D^{0*} \bar{D^0} + c.c., \gamma J/\psi, \pi^+\pi^- J/\psi, \gamma \psi(3686), \omega J/\psi$, etc. decay modes are all observed in the Belle, BABAR, BESIII, and LHCB experiments. $X(3872) \to D^{*0}\bar {D^{0}}+ c.c. $ decay of branching ratio is about 52.4\% ~\cite{X3872-DD}. Therefore, the total yield of $X(3872)$ generated in the high-energy $pp$ collision is approximately twice that of our simulation results.

Fig.2 shows the transverse momentum $p_{T}$ distributions of the $X(3872)$ as different structures using different bound states in $pp$ collision at $\sqrt{s}=7$ and 13 TeV. Obviously, the shape of the $p_T$ distribution of $X(3872)$ produced by different decay final states and different c.m. energies are similar to each other. But the molecular state $X^m(3872)$ has a wider $p_T$ distribution than tetraquark state $X^t(3872)$ and nucleus-like state $X^n(3872)$.

We also predicted the rapidity distribution of $X(3872)$ are also considered as above three structures in $pp$ collisions at $\sqrt{s}=7$ and 13 TeV, which are symmetrically distributed in the range from $-7.5$ to 7.5, shown in Fig. 3. From this figure, one can see that the rapidity distributions for the three $X(3872)$ structures are quite similar to each other except that for the molecular state is much narrower compared to the others.

\section{conclusions}

In this paper, $D^{0}$, $D^{+}$, $D^{*+}$, $\pi^{+}$ and $K^{+}$ are generated in $pp$ collision at $\sqrt{s}=7$ and 13 TeV by using the PACIAE model. The productions of each particle are consistent with the experimental data.
 Then the $D$ and $\bar D^{*}$ are put into the DCPC model to construct the clusters of $D^{*0}\bar{D}^{0}$, $D^{0}\bar{D}^{*0}$, $D^{-}{D}^{*+}$ and $D^{+}{D}^{*-}$ to produce $X(3872)$. The production and the characteristics of the $X(3872)$ as tetraquark state, nucleus-like state, molecular state are predicted using different bound states. We found that the yields of $X(3872$) in $pp$ collisions mainly comes from the $D^{*0}\bar{D}^{0}$ and $D^{0}\bar{D}^{*0}$ clusters. If we take half of its decay width as the mass uncertainty $\Delta m$, the yields of the $X(3872)$ as tetraquark state, nucleus-like state, and molecular state in $pp$ collision are predicted to be $8.22\times10^{-8}$, $4.21\times10^{-7}$, $1.01\times10^{-6}$ at $\sqrt{s}=7$ TeV and $9.00\times10^{-8}$, $4.72\times10^{-7}$, $1.18\times10^{-6}$ at $\sqrt{s}=13$ TeV, respectively. The transverse momentum $p_{T}$ distribution of $X(3872)$ as tetraquark state, nucleus-like state are similar to each other while the rapidity distribution for the molecular state of $X(3872)$ is slightly wider than that of other two structures.

To further understand the nature of exotic state $X(3872)$, we therefore suggest to measure their yields and properties in $pp$ and heavy-ion collisions by the ALICE and STAR experiments.

\vspace{4mm}

\textbf{Acknowledgments}
The work of G. Chen is supported by the National Natural Science Foundation of China (NSFC) (11475149), of D. M. Zhou is supported by the NSFC (11705167), of L. Zheng is supported by the NSFC (11905188) and the Innovation Fund of  of Key Laboratory of Quark and Leption Physics QLPL2020P01, and of X. L. Kang is supported by the NSFC (12005195).


\vspace{4mm}
\textbf{Data Availability Statement} This manuscript has no associated data
or the data will not be deposited. [Authors' comment: All data generated
during this study are contained in this published article.]

\end{document}